\documentclass[aps,prl,reprint,onecolumn,12pt,amsmath,amssymb,citeautoscript,superscriptaddress,nobibnotes]{revtex4-1}

\usepackage{bm,mathrsfs,dcolumn,graphicx,color}
\usepackage{amsmath}
\usepackage{graphicx}
\usepackage[T1]{fontenc}
\usepackage{hyphenat}
\usepackage{setspace}
\usepackage{setspace}
\pdfoutput=1

\newcommand{\red}[1]{{\color[rgb]{0,0,0}{#1}}} 
\newcommand{\blue}[1]{{\color[rgb]{0,0,0}{#1}}} 
\newcommand{\green}[1]{{\color[rgb]{0,0,0}{#1}}} 
\newcommand{\violet}[1]{{\color[rgb]{0,0,0}{#1}}} 
\newcommand{\rust}[1]{{\color[rgb]{0,0,0}{#1}}}

\newcommand{\arr}[1]{{\color[rgb]{0,0,0}{#1}}} 

\newcommand{\ac}[1]{{\color[rgb]{0,0,0}{#1}}}

\newcommand{\new}[1]{{\color[rgb]{0,0,0}{#1}}}

\begin{document}

\title{Coulomb engineering of the bandgap in 2D semiconductors} 

\author{Archana Raja\footnote{archana.raja@stanford.edu}}
\affiliation{Departments of Physics and Electrical Engineering, Columbia University, New York, New York 10027, USA}
\affiliation{Department of Chemistry, Columbia University, New York, New York 10027, USA}
\affiliation{Department of Applied Physics, Stanford University, Stanford, California 94305, USA}
\affiliation{SLAC National Accelerator Laboratory, Menlo Park, California 94025, USA}
\author{Andrey Chaves}
\affiliation{Department of Chemistry, Columbia University, New York, New York 10027, USA}
\affiliation{Universidade Federal do Cear\'a, Departamento de F\'isica, Caixa Postal 6030, Fortaleza 60455-760, Cear\'a, Brazil}
\author{Jaeeun Yu}
\affiliation{Department of Chemistry, Columbia University, New York, New York 10027, USA}
\author{Ghidewon Arefe}
\affiliation{Department of Mechanical Engineering, Columbia University, New York, New York 10027, USA}
\author{Heather M. Hill}
\author{Albert F. Rigosi}
\affiliation{Department of Applied Physics, Stanford University, Stanford, California 94305, USA}
\affiliation{Departments of Physics and Electrical Engineering, Columbia University, New York, New York 10027, USA}
\author{Timothy C. Berkelbach}
\affiliation{Department of Chemistry and James Franck Institute, University of Chicago, Chicago, Illinois 60637, USA}
\author{Philipp Nagler}
\author{Christian Sch\"uller}
\author{Tobias Korn}
\affiliation{Department of Physics, University of Regensburg, Regensburg D-93040, Germany}
\author{Colin Nuckolls}
\affiliation{Department of Chemistry, Columbia University, New York, New York 10027, USA}
\author{James Hone}
\affiliation{Department of Mechanical Engineering, Columbia University, New York, New York 10027, USA}
\author{Louis E. Brus}
\affiliation{Department of Chemistry, Columbia University, New York, New York 10027, USA}
\author{Tony F. Heinz}
\affiliation{Department of Applied Physics, Stanford University, Stanford, California 94305, USA}
\affiliation{SLAC National Accelerator Laboratory, Menlo Park, California 94025, USA}
\affiliation{Departments of Physics and Electrical Engineering, Columbia University, New York, New York 10027, USA}
\author{David R. Reichman}
\affiliation{Department of Chemistry, Columbia University, New York, New York 10027, USA}
\author{Alexey Chernikov\footnote{alexey.chernikov@ur.de}}
\affiliation{Departments of Physics and Electrical Engineering, Columbia University, New York, New York 10027, USA}
\affiliation{Department of Physics, University of Regensburg, Regensburg D-93040, Germany}

\begin{abstract}
The ability to control the size of the electronic bandgap is an integral part of solid-state technology. 
Atomically-thin two-dimensional crystals offer a new approach for tuning the energies of the electronic states based on the interplay between the environmental sensitivity and  unusual strength of the Coulomb interaction in these materials. By engineering the surrounding dielectric environment, we are able to tune the electronic bandgap in monolayers of WS$_2$ and WSe$_2$ by hundreds of meV. We exploit this behavior to present an \textit{in-plane} dielectric heterostructure with a spatially dependent bandgap, illustrating the feasibility of our approach \new{for the creation of  lateral junctions with nanoscale resolution.} 
This successful demonstration of bandgap engineering based on the non-invasive modification of the Coulomb interaction should enable the design of a new class of atomically thin devices to advance the limits of size and functionality for solid-state technologies.
\end{abstract}
\maketitle

The precise and efficient manipulation of electrons in solid-state devices has driven remarkable progress across \arr{fields from information processing and communication technology to sensing and renewable energy.}
The ability to engineer the electronic bandgap, the forbidden energy region separating occupied and unoccupied electronic states, is crucial to these applications\,\cite{Capasso1987}. 
Several methods currently exist to tune a material's bandgap by altering, for example, its chemical composition, spatial extent (quantum confinement), background doping, or lattice constant via mechanical strain\,\cite{Klingshirn2007}. 
\new{Such methods are, however, perturbative in nature and not suitable for making arbitrarily shaped, atomically sharp variations in the bandgap without degrading the intrinsic properties of the material. Consequently, there is a need to approach this important problem from a fresh perspective.}

The emerging class of atomically-thin two-dimensional (2D) materials derived from bulk van der Waals crystals offers an alternative route \red{to bandgap engineering}. 
Within the family of 2D materials, much recent research has focused on the semiconducting transition-metal dichalcogenides (TMDCs) - MX$_2$ with M\,$=$\,Mo,\,W and X\,$=$\,S,\,Se,\,Te\,\cite{Novoselov2005}. 
In the monolayer limit, \arr{these} TMDCs are direct-gap semiconductors with optical gap in the visible and near-IR spectral range\,\cite{Splendiani2010,Mak2010}.
They combine strong inter- and intraband light-matter coupling\,\cite{Zhang2014a,Poellmann2015} with intriguing spin-valley physics\,\cite{Xu2014,Yu2015,Stier2016}, high charge carrier mobilities\,\cite{Jariwala2014,Cui2015}, ready modification of the \textit{in-plane} material structure~\cite{Gong2014,Huang2014,Kappera2014,Guo2015}, and seamless integration into a variety of van der Waals heterostructures\,\cite{Geim2013}.

\begin{figure}
	\singlespacing
	\footnotesize
	\centering
		\includegraphics[width=16 cm]{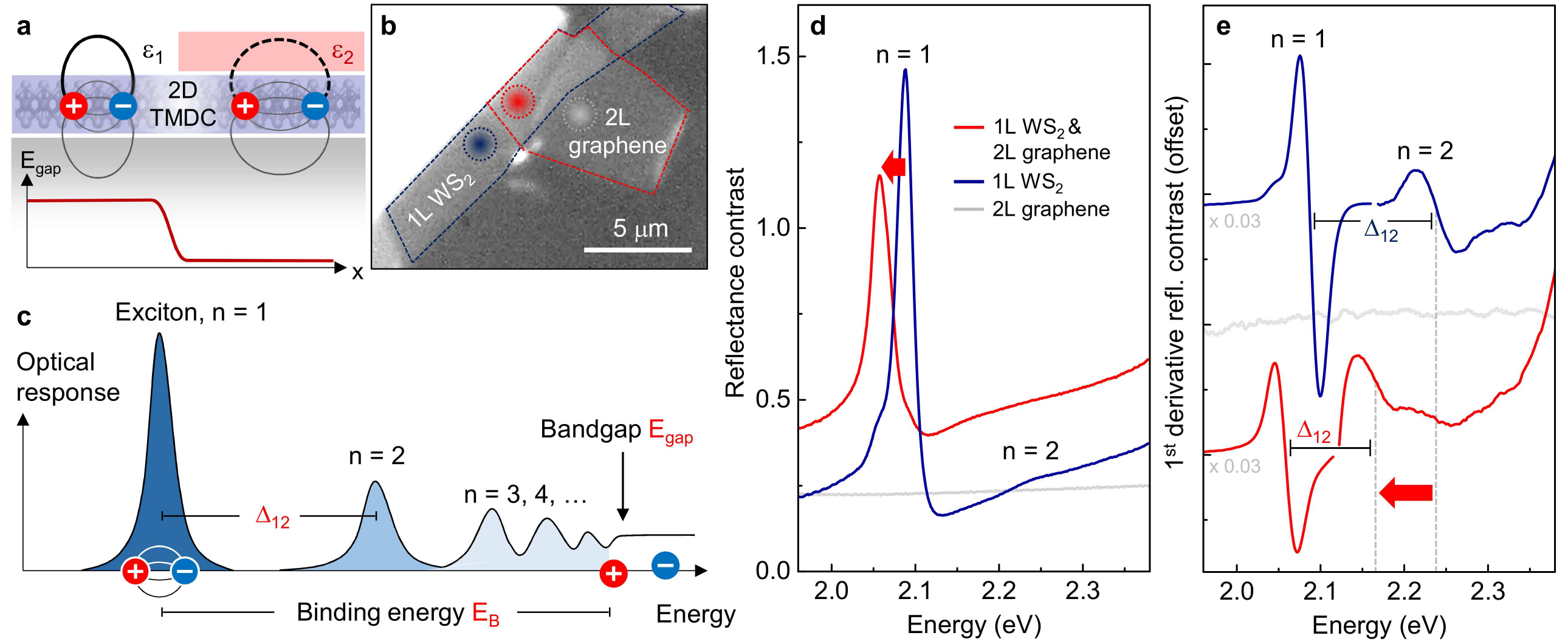}
		\caption{\textbf{Engineering Coulomb interactions through environmental screening.} 
            \textbf{a} Schematic illustration of a semiconducting 2D TMDC material, partially covered with an ultra-thin dielectric layer. The strong Coulomb interaction between charged particles in low-dimensional systems affects both the exciton binding energy and the quasiparticle bandgap. The interaction can be strongly modified by modulating the environmental dielectric screening on atomic length scales.
								\textbf{b} An optical micrograph of the heterostructure under study: monolayer WS$_2$ covered with a bilayer of graphene. Dotted circles indicate  positions for the optical measurements.
                \textbf{c} Illustration of the optical response of an ideal 2D semiconductor, including exciton ground and excited state resonances and the onset of the (quasiparticle) bandgap.  
								\textbf{d} Reflectance contrast spectra of the bare bilayer graphene, monolayer WS$_2$, and the resulting WS$_2$ / graphene heterostructure at a temperature of 70\,K.  						
								\blue{\textbf{e} First derivatives of the reflectance contrast spectra in \textbf{d} (after averaging over a 20\,meV interval), offset for clarity.
								Peak positions of the exciton ground state ($n=1$) and the first excited state ($n=2$) resonances, roughly corresponding to the points of inflection, are indicated by dashed lines; $\Delta_{12}$ denotes the respective energy separations.}								 
								The observed decrease of $\Delta_{12}$ across the in-plane boundary of the heterostructure is indicative of a reduction \arr{of} the exciton binding energy and bandgap by more than 100\,meV due to the presence of the adjacent graphene bilayer. 
			}
\label{fig1}
\end{figure}

Importantly, the Coulomb interactions between charge carriers in atomically thin TMDCs are remarkably strong\,\cite{Cheiwchanchamnangij2012,Ramasubramaniam2012,Qiu2013,Berkelbach2013}.
\arr{This leads to a significant renormalization of the electronic energy levels and increase in size of the quasiparticle bandgap.}
\ac{The Coulomb interactions are also reflected in the binding energies of excitons, i.e., tightly bound electron-hole pairs\,\cite{Klingshirn2007}, that are more than an order of magnitude greater in TMDC monolayers than in typical inorganic semiconductors~\cite{Zhang,He2014,Chernikov2014,Ye2014,Ugeda2014}.} 
The strength of the Coulomb interaction in these materials originates from weak dielectric screening in the two-dimensional limit~\cite{Keldysh1979,Cudazzo2011,Berkelbach2013}.  
For distances on the order of a few nanometers or greater, the screening is determined by the immediate surroundings of the material, which can be vacuum or air in the ideal case of suspended samples.  
More generally, the interaction between charge carriers is highly sensitive to the local dielectric environment\,\cite{Keldysh1979,Cudazzo2011,He2014,Chernikov2014,Ugeda2014,Lin2014}.
\arr{Correspondingly}, both the electronic bandgap and the exciton binding energy are expected to be highly tunable \arr{by means of} a deliberate change of \blue{this} environment, as illustrated in Fig.\,\ref{fig1}a, \arr{like} the influence of a solvent on the properties of molecules, quantum dots, carbon nanotubes, and other nanostructures suspended in solution\blue{\,\cite{Brus2014,Walsh2007,Malapanis2011}}.  
In addition, the passivated and chemically inert van der Waals surface allows for several atomically-thin layers to be brought into close proximity while \arr{still} retaining the intrinsic properties and functionality of the individual components\,\cite{Geim2013}. 
These observations motivate a unique program to explore the concept of \green{``Coulomb engineering" of the bandgap} \arr{by} local changes in the dielectric environment. 
This strategy offers a new and non-invasive means of locally tuning the energies of the electronic states \arr{in 2D materials}, even allowing in-plane heterostructures down to nanometer length scales~\cite{Rosner2016}.  
\new{As a result, it not only effectively demonstrates the validity of fundamental physics with respect to the Coulomb interaction in atomically-thin systems, but offers a viable opportunity to directly harness many-body phenomena for future technology.}

In this report, we provide direct experimental demonstration of control of the bandgap in a 2D semiconductor using Coulomb engineering of the local dielectric environment.
By placing layers of graphene above and below monolayers (1L) of WS$_2$ and WSe$_2$, we achieve tuning of the electronic quasiparticle bandgap, as well as of the exciton binding energy of the two TMDC monolayers by several 100's of meV.
\new{We note that graphene is particularly well-suited to demonstrate and explore the concept of dielectric heterostructures.
It ideally combines a high dielectric screening constant with the possibility to add an arbitrary number of additional layers as thin as only 3\,\AA.
Furthermore, the TMDC/graphene structures have been heavily studied recently in a variety of contexts with potential applications in optoelectronics and photovoltaics\,\cite{Roy2013,Georgiou2013,Bertolazzi2013,Bernardi2013}.}
Screening is found to be maximized for just a few layers of graphene as the surrounding dielectric, showing that Coulomb\blue{-}engineered bandgaps can be \green{realized} with a spatial-resolution of around a nanometer.
Moreover, an \textit{in-plane} heterostructure with a spatially-dependent electronic bandgap is shown to exhibit a potential well on the order of more than 100\,meV, illustrating the feasibility of our approach for applications under ambient and even high-temperature conditions.  
Our results are supported by theoretical calculations employing a quantum mechanical Wannier exciton model~\cite{Berkelbach2013}. 
The dielectric screening leading to the bandgap renormalization can be treated in a semiclassical electrostatic framework that accounts for the underlying substrate and the additional graphene layers.  
For a more quantitative description of the screening, we employ a recently developed quantum electrostatic heterostructure approach from Ref.\,\onlinecite{Andersen2015}.   

An optical micrograph \arr{of} a typical \blue{sample,} 1L\,WS$_2$ partially covered with bilayer\,(2L) graphene\blue{,} is presented in Fig.\,\ref{fig1}b.  
To monitor the quasiparticle bandgap of the material, \red{we \blue{first} identify the energies of the excitonic resonances in different dielectric environments using optical reflectance spectroscopy.} 
The optical response of an ultra-thin 2D semiconductor is illustrated schematically in Fig.\,\ref{fig1}c. 
The Coulomb attraction between electrons and holes leads to the emergence of bound exciton states below the quasiparticle bandgap\,\cite{Klingshirn2007,Haug2009,Kazimierczuk2014}, which are labeled according to their principal quantum number $n=1, 2, 3,...$, analogous to the \arr{states of the} hydrogen atom.
\blue{(Throughout the rest of the manuscript \violet{we omit} the term \violet{"quasiparticle" for} clarity of presentation.)}
The difference between the bandgap $E_{\mathrm{gap}}$ and the exciton resonance energies defines the respective exciton binding energies.
In particular, the energy $\Delta_{12}$ between the exciton ground state ($n=1$) and the first excited state ($n=2$) scales with the ground state exciton binding energy $E_B$.
This allows us to determine the \blue{size of the} bandgap from the transition energy $E_{1}$ of the exciton ground state via $E_{\mathrm{gap}} = E_{1} + E_B$.

\green{Typical} linear reflectance contrast spectra, $\Delta R/R=(R_{\rm sample}-R_{\rm substrate})/R_{\rm substrate}$, of the bare 2L~graphene, 1L~WS$_2$, and the resulting heterostructure \blue{at T\,=\,70\,K} are presented in Fig.\,\ref{fig1}d. 
For \blue{such} ultrathin layers with moderate reflectance contrast signals on transparent substrates, the quantity $\Delta R/R$ is predominantly determined by the imaginary part of the dielectric function, which is proportional to the optical absorption\,\cite{Mak2008,Li2014}.
In the spectral region shown, the \arr{response of} 1L\,WS$_2$ is dominated by the creation of so-called $A$ excitons at the fundamental optical transition in the material, at the $K$ and $K^\prime$ points of the hexagonal Brillouin zone. 
In particular, the ground-state ($n=1$) excitonic resonance occurs at 2.089\,eV.  
The first excited state $n\,=\,2$ appears as a smaller spectral feature at 2.234\,eV, with an energy separation between the two states of $\Delta_{12}$\,=\,156\,meV.
\blue{In addition, the first derivatives of $\Delta R/R$ are presented in Fig.\,\ref{fig1}e, where the spectral region in the range of the $n=1$ state is scaled by factor 0.03 for better comparison.
Here, the \arr{energies} of the peaks correspond to the points of inflection of the asymmetric derivative features, as indicated by dashed lines for the $n=2$ states.}
\blue{Finally,} the shoulder on the low-energy side of the $n=1$ peak at 2.045\,eV \arr{arises from} charged excitons, indicating \arr{slight} residual doping in the WSe$_2$ material\,\cite{Chernikov2015b,Plechinger2015}. 
\blue{Overall,} the \blue{1L WS$_2$ response} matches our previous observations on uncapped \blue{sample}\green{s supported on} fused silica\,\cite{Chernikov2014,Chernikov2015b}, consistent with an exciton binding energy on the order of 300\,meV.
For bilayer graphene, we recover the characteristic flat \blue{reflectance} \green{contrast} \arr{over the relevant} spectral range\,\cite{Mak2011}.

In case of WS$_2$ capped with graphene, the overall reflectance contrast is offset by the graphene reflectance, similar to findings in TMDC/TMDC heterostructures \,\cite{Rigosi2015}.  
Most importantly, however, we observe pronounced shifts of the WS$_2$ exciton resonances to lower energies, where the $n\,=\,1$ transition and the $n\,=\,2$ states are now located at 2.060 and 2.167\,eV, respectively (see Figs.\,\ref{fig1}d and \blue{e}). 
The corresponding decrease of $\Delta_{12}$ from 156 to 107\,meV is indicative of a strong reduction in the exciton binding energy and bandgap. 
In particular, the absolute shift of the $n\,=\,2$ state by almost 70\,meV \arr{defines} the minimum expected decrease in the bandgap.
More quantitatively, by assuming a similar \blue{non-hydrogenic} scaling \arr{like that} in \blue{Ref.}\,\cite{Chernikov2014}, \blue{i.e., $E_B=2\Delta_{12}$,} the reduction in exciton binding energy is estimated to be on the order of 100\,meV, \blue{from 312\,meV in bare WS$_2$ to 214\,meV} in WS$_2$ capped by \blue{2L} graphene.  
\arr{From} $E_{\mathrm{gap}} = E_{1} + E_B$, \arr{we infer a} bandgap for bare WS$_2$ \arr{of} 2.40\,eV, \arr{reducing} to 2.27\,eV in the WS$_2$/graphene heterostructure.
\arr{We thus see} a 130\,meV decrease in the bandgap energy \arr{from} the presence of the capping layer.

\begin{figure}[t]
	\centering
		\includegraphics[width=6.5 cm]{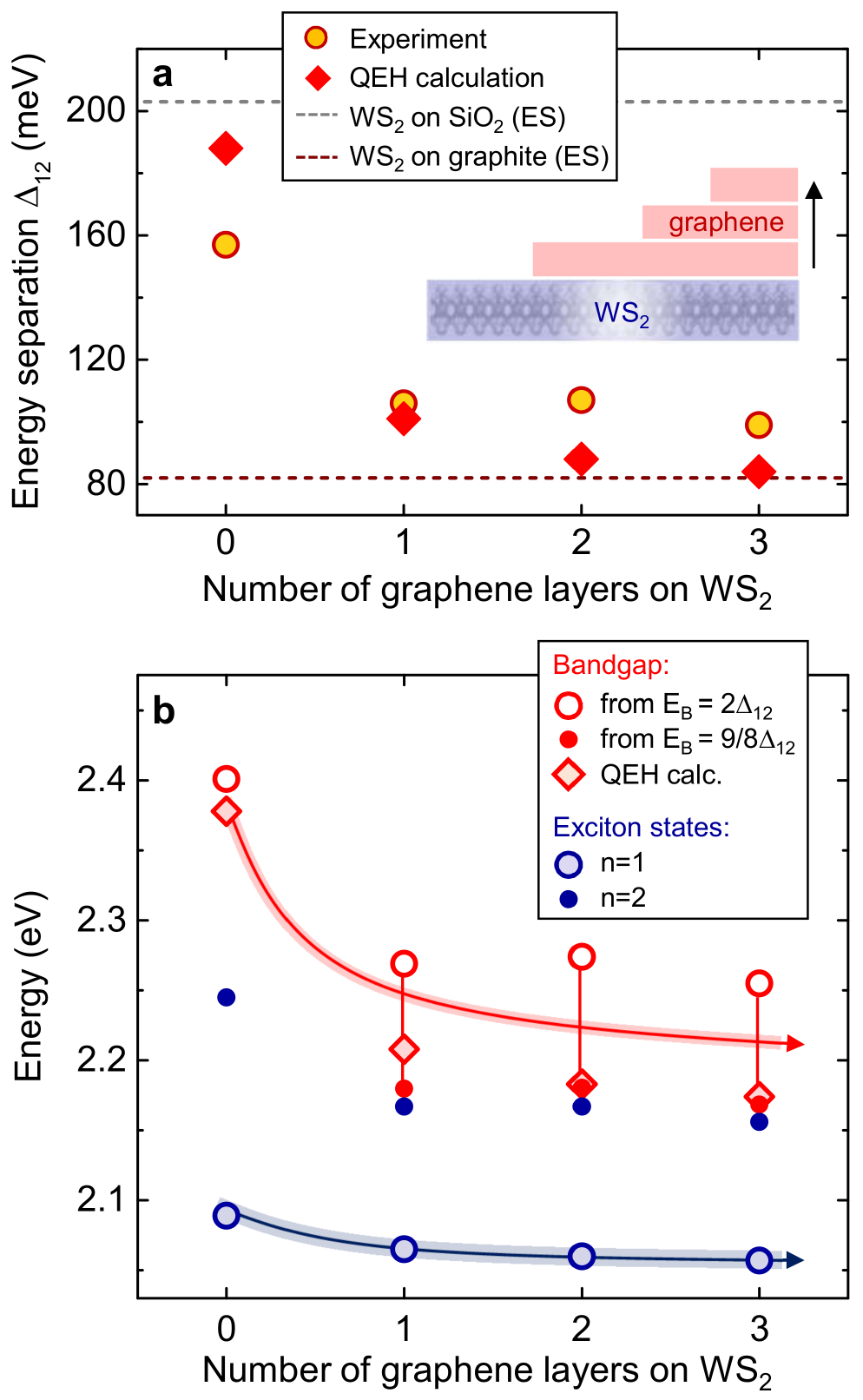}
		\caption{\textbf{Tuning exciton and bandgap energies in 1L WS$_2$ by capping with graphene layers.} 
                \textbf{a} Experimentally and theoretically obtained energy separation $\Delta_{12}$ between the $n=1$ and $n=2$ exciton states as a function of the number of layers of capping graphene. Dashed lines indicate $\Delta_{12}$ values from the solution of the electrostatic (ES) model for uncapped WS$_2$ supported by fused silica substrate (grey) and covered with \textit{bulk} graphite (red), representing two \arr{ideal limiting} cases.
                \blue{\textbf{b}\,\,Absolute energies of the experimentally measured exciton ground state ($n=1$) and the first excited state ($n=2$) resonances, as well as the estimated positions of the bandgap obtained by adding the exciton binding energy to the energy of the $n=1$ state.
								The binding energy scales with $\Delta_{12}$, where the limiting cases are an experimentally determined non-hydrogenic scaling for WS$_2$ on SiO$_2$ substrate from Ref.\,\cite{Chernikov2014} ($E_B = 2\Delta_{12}$) and the 2D-hydrogen model in a homogeneous dielectric ($E_B = 9/8\Delta_{12}$).
								These are compared to the bandgap \arr{energies deduced} from the calculated exciton binding energies using the QEH model.
								The solid lines are guides to the eye.}
}
\label{fig2}
\end{figure}

To understand these \arr{experimental} findings \arr{more intuitively}, we recall that although the excitons are confined to the WS$_2$ layer, the electric field between the constituent electrons and holes permeates both the material and the local \blue{surroundings} (Fig.\,\ref{fig1}a). 
In particular, the screening \arr{for larger electron-hole separations} is increasingly dominated by the dielectric properties of the environment.  
Therefore, the strength of the Coulomb interaction is reduced by the addition of graphene layers on top of WS$_2$, \red{leading to a decrease in} \blue{both} the exciton binding energy and the bandgap.
A critical remaining question concerns the degree of spatial locality of the modulation of the electronic structure induced by the dielectric environment. We have been able to address this issue spectroscopically with sub-nanometer precision.  To do so,  we simply track the change in the WS2 bandgap for dielectric screening when the semiconductor is capped by 1, 2, or 3 layer graphene.
The extracted exciton peak separation energy $\Delta_{12}$ and the corresponding evolution of the bandgap are presented in Figs.\,\ref{fig2}a and \,\ref{fig2}b, respectively. 
\ac{Remarkably,} we observe the strongest change \ac{already} \arr{from} the first graphene layer, which is followed by rapid saturation with increasing thickness within experimental uncertainty.
This \arr{result strongly} suggests that the change in bandgap should also \arr{occur on a \ac{similar ultra-short} length scale} at the \emph{in-plane} boundary of the uncapped and graphene-capped WS$_2$, \blue{consistent with predictions from Ref.\,\cite{Rosner2016}}.

For a more precise analysis of our findings we turn to a Wannier-like exciton model~\cite{Berkelbach2013}, for which the environment enters through the \red{non-local} screening of the electron-hole Coulomb interaction.
To handle atomistically complex dielectric environments, we employ the recently introduced quantum electrostatic heterostructure \blue{(QEH)} approach presented in Ref.\,\cite{Andersen2015}. 
Within this model, the electrostatic potential between electrons and holes confined to a 2D layer can be obtained for nearly arbitrary vertical heterostructures, taking into account the precise alignment of the individual materials and the resulting spatially dependent dielectric response. 
The exciton states are subsequently calculated by solving the Wannier equation in the effective mass approximation with an exciton reduced mass of 0.16\,$m_0$ as obtained from \textit{ab initio} calculations\,\cite{Berkelbach2013}.
\ac{To account for the dielectric screening from the environment, mainly through the underlying fused silica substrate and potential adsorbates such as water, we adjust the effective dielectric constant below the 2D layer, resulting in $\Delta_{12}=188\,meV$ and $E_B=289\,meV$, roughly matching experimental observations.} 
Then, additional graphene layers are added on top of the WS$_2$ \arr{monolayer} with all parameters being fixed.  
The interlayer separation between WS$_2$ and graphene is set to 0.5\,nm, corresponding to the average of the interlayer separations for the materials in literature~\cite{Gutierrez2012,Baskin1955}.

\begin{figure}[ht]
	\centering
		\includegraphics[width=8.4 cm]{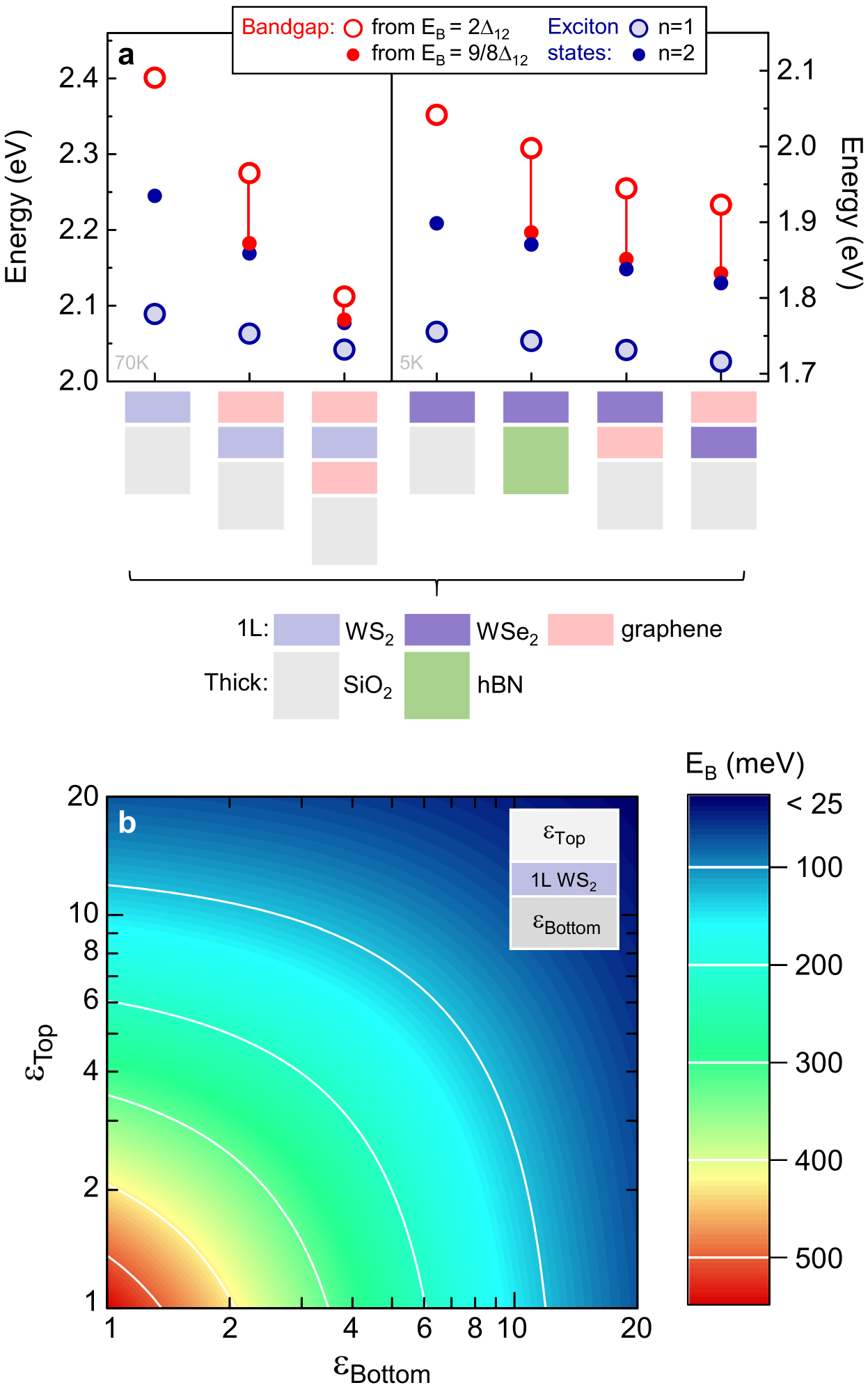}
		\caption{\blue{\textbf{Influence of \arr{the choice of material} on the dielectric tuning \green{of the bandgap}.} 
                \textbf{a} Experimentally measured exciton ground state ($n=1$) and the first excited state ($n=2$) transition energies, as well as the estimated shifts of the bandgap for a variety of heterostructures.
							Their respective stacking compositions are indicated along the horizontal axis.
							The bandgap is obtained by adding the exciton binding energy to the \arr{measured transition} energy of the $n=1$ state.
							To estimate the binding energy from the energy separation of the exciton states $\Delta_{12}$, we considered the limiting cases of a non-hydrogenic scaling from Ref.\,\cite{Chernikov2014} ($E_B = 2\Delta_{12}$) and the 2D-hydrogen model ($E_B = 9/8\Delta_{12}$).
							\textbf{b} An overview of predicted changes in the exciton binding energy in 1L WS$_2$, encapsulated between two thick layers of dielectrics.
							The binding energy $E_B$ is calculated by using \arr{the} electrostatic approach in \arr{the} effective mass approximation and presented in a 2D false-color plot as \arr{a} function of the top and bottom dielectric constants.
							The changes in the magnitude of $E_B$ are roughly equal to the corresponding shifts of the bandgap \arr{and can reach 500\,meV}}
}
\label{fig3}
\end{figure}
The theoretically predicted energy separation $\Delta_{12}$ is plotted in Fig.~\ref{fig2}a as a function of the number of graphene layers and compared to experiment. 
The calculations reproduce both the abrupt change and the subsequent saturation of $\Delta_{12}$ with graphene thickness. 
Furthermore, the absolute energy values are in semi-quantitative agreement with \violet{the} measurements, supporting the attribution of the measured change of $\Delta_{12}$ to the dielectric screening from adjacent graphene layers.
The model also agrees with a classical electrostatic screening theory for the limiting cases of an uncapped WS$_2$ monolayer on fused silica and for a layer fully covered with bulk graphite on top, the results of which are indicated by dashed lines in Fig.\,\ref{fig2}a (see Supplementary Information for details).  
The calculated exciton binding energy changes from 290\,meV for uncapped WS$_2$ to 120\,meV for the case of a trilayer graphene heterostructure.
As previously discussed, the binding energies together with the absolute energies of the exciton ground state resonances can be used to infer the \blue{size of the} bandgap. 
\blue{The evolution of the bandgap and the corresponding $n=1$ \blue{and $n=2$} exciton transition energies are presented in Fig.\,\ref{fig2}b.
The binding energies obtained from the QEH model are compared with experimentally
determined limits from the relation $E_B \propto \Delta_{12}$ by assuming a non-hydrogenic scaling $E_B = 2\Delta_{12}$ as was observed for a single WS$_2$ layer on
SiO$_2$\,\cite{Chernikov2014} or conventional 2D hydrogenic scaling with $E_B =
9/8\Delta_{12}$ \arr{for an} homogeneous dielectric.
These two relations provide\arr{, respectively,} boundaries for the scaling in generic heterostructures of 1L TMDCs embedded in \arr{a} dielectric environment with higher dielectric screening than the SiO$_2$ support and lower dielectric screening than the corresponding bulk crystals.
In general, the scaling of $E_B$ with $\Delta_{12}$ converges towards the 2D-hydrogen model as the screening of the surroundings approachesthe screening constant that of the bulk TMDC.
For the case of trilayer graphene, this simple estimate implies a bandgap reduction of at least 150\,meV and at most 230\,meV.}

\begin{figure}[ht]
	\centering
		\includegraphics[width=8.6 cm]{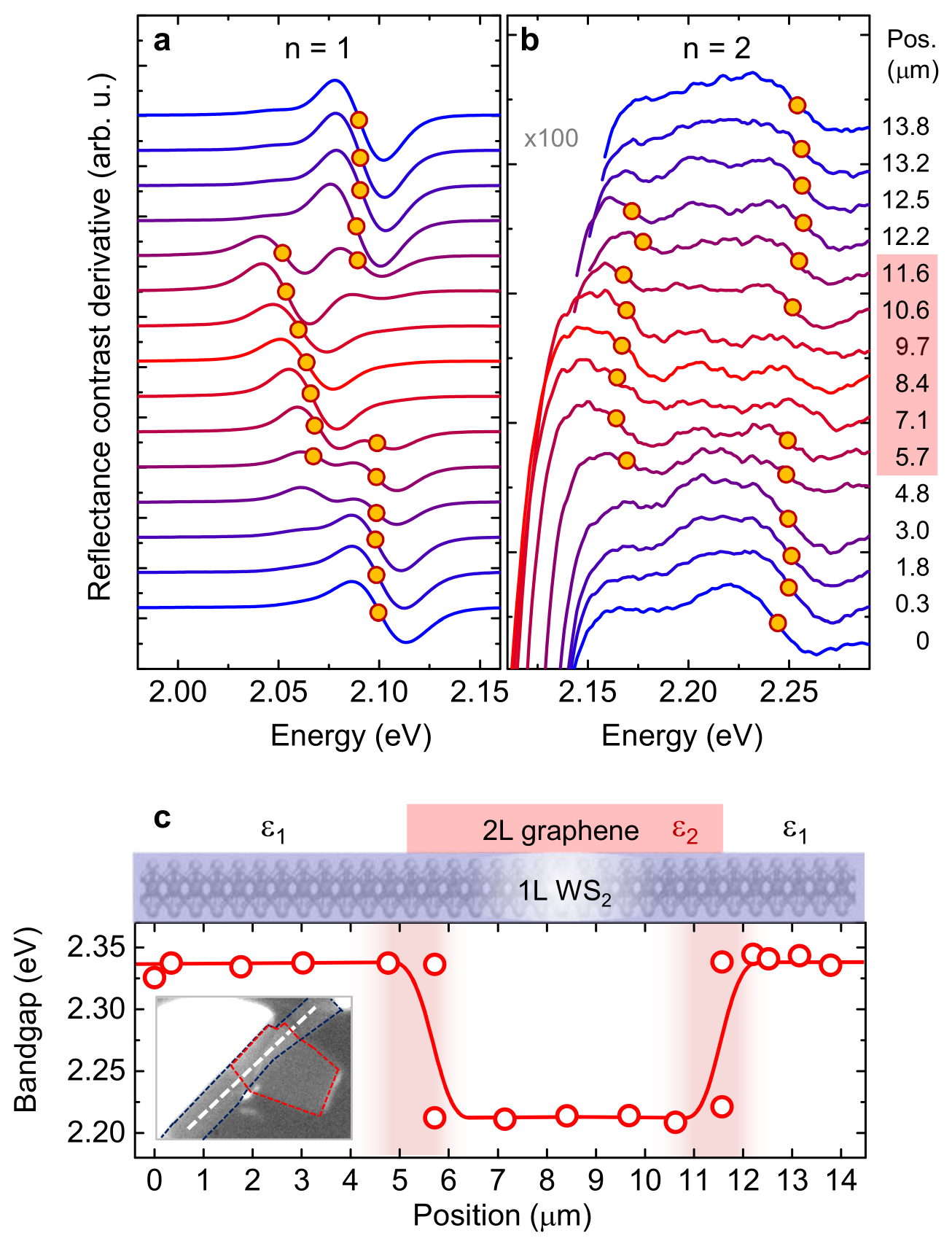}
		\caption{\textbf{In-plane heterostructure via Coulomb engineering of monolayer WS$_2$.} 
                \textbf{a} First-order derivatives of the reflectance contrast of a 1L\,WS$_2$ sample for varying spatial positions across the lateral 1L\,WS$_2$\,/\,2L~graphene boundary. The data are shown in the spectral range of the exciton ground state ($n=1$) resonance and vertically offset for clarity.
                \textbf{b} For the spectral range of the excited state ($n=2$) of the exciton with the vertical axis scaled by factor of 100 for direct comparison. Full circles in \textbf{a} and \textbf{b} indicate peak energies of the resonances, corresponding to the points of inflection of the derivatives.
                \textbf{c} Spatially-dependent bandgap energy extracted from the exciton peak positions along the profile of the lateral WS$_2$/graphene heterostructure, as illustrated in the schematic representation and marked by the dashed line in the optical micrograph. The shaded areas mark fundamental indicate the diffraction limit for the spatial resolution in the optical setup and the solid line is a guide to the eye.
	}
\label{fig4}
\end{figure} 

\blue{In addition to the graphene-capped WS$_2$ samples, a variety of heterostructures was investigated in a similar manner.  
These include 1L WS$_2$ encapsulated between two graphene layers, graphene-capped 1L WSe$_2$, graphene-supported 1L WSe$_2$, and 1L WSe$_2$ on an 8\,nm layer of hexagonal boron nitride (hBN).
In all cases, a decrease in $\Delta_{12}$ separation was observed with increasing dielectric screening of the environment (see Supplementary Information for individual reflectance spectra and additional sample details).  
A summary of the results is presented in Fig.\,\ref{fig3}a, including experimentally obtained $n=1$ and $n=2$ transition energies, as well as the corresponding shifts of the bandgap, estimated as above.  
The bandgap of WSe$_2$ can be thus tuned by more than 100\,meV and the largest shift of almost 300\,meV is observed for graphene-encapsulated WS$_2$, the structure with the highest dielectric screening.
For comparison, the influence of a generic an arbitrary dielectric environment is presented in Fig.\,\ref{fig3}b, which shows the calculated exciton binding energy of 1L WS$_2$ encapsulated between two thick layers of varying dielectric constants.  
As we have shown, the change in the bandgap is roughly the same as the change in the binding energy and thus can be as high as 500\,meV (corresponding to the intrinsic value of the exciton binding energy for a suspended sample).
}

Finally, we demonstrate an \textit{in-plane} \arr{2D semiconductor} heterostructure with a spatially-dependent bandgap \blue{profile}. 
We have produced an important building block for future devices by constructing a spatially varying dielectric environment surrounding the semiconductor.
We analyze the response using spatially resolved optical measurements, corresponding to the line-map shown in Fig.\,\ref{fig4}c and in the inset.
Using spatially-resolved measurements, we scan across the structure (cf. Fig.\,\ref{fig1}b) through the regions of bare WS$_2$ and WS$_2$ covered by a bilayer of graphene.
The corresponding path is illustrated schematically in Fig.\,\ref{fig4}c and in the inset.
First-order derivatives of the reflectance contrast spectra are presented in Figs.\,\ref{fig4}a and b in the spectral range of the WS$_2$ exciton $n=1$ and $n=2$ resonances, respectively.  
Each spectral trace corresponds to a different spatial position $x$ on the sample; the bilayer graphene flake covers the WS$_2$ monolayer between $5\,\mu m$ and $12\,\mu m$ on the x-axis.
Like the data shown in Figs.\,\ref{fig1}d \violet{and 1e}, both the ground and excited state resonances of the WS$_2$ excitons shift to lower energies in the presence of graphene. 
The peak energies are extracted from the points of inflection of the derivative, indicated by circles in Figs.\,\ref{fig4}a and \ref{fig4}b.
The appearance of multiple transitions in the same spectrum \arr{reflects the limited} spatial resolution ($1\,\mu m$), as well as due to a small amount of the WS$_2$ \arr{monolayer} not being in close contact with graphene (see Supplementary Information for details).  

The spatial dependence is presented in Fig.\,\ref{fig4}c along the path marked in the optical micrograph (inset), which includes two WS$_2$/graphene in-plane junctions.  
As previously discussed, the induced energy shifts result in an overall decrease of the relative energy separation $\Delta_{12}$ from \blue{about} 160\,\rust{meV,} \blue{down} to 105\,meV.  
\violet{Here,} the binding energy is extracted by multiplying $\Delta_{12}$ with the scaling factor deduced from the QEH calculations presented in Fig.\,\ref{fig2} (1.54 and 1.40 for the bare and 2L~graphene-covered sample respectively) to obtain the bandgap at each point.
The resulting bandgap profile is representative of a potential well (\green{graphene-}covered area) surrounded by two adjacent barriers at higher energies (bare sample).
Model self-energy calculations on monolayer TMDCs in structured dielectric environments~\cite{Rosner2016} suggest that the \red{interface between the uncapped and capped regions} should yield an in-plane type-II heterostructure. 
In particular, the areas capped by graphene \red{are expected to} have a higher local valence band that acts as a potential well for holes\rust{.}
\rust{The dielectric effect on the conduction band is predicted to be weaker, with a slightly higher energy for the capped regions leading to a small barrier for electron flow from the bare to capped regions.}
Since the overall energy shifts of the bandgap are larger than thermal energy at room temperature, our results render the observed phenomenon technologically promising for applications under ambient or even high-temperature conditions.

In conclusion, we have demonstrated a new approach to the engineering of electronic properties through local dielectric screening of the Coulomb interaction in 2D heterostructures.  
We have shown tuning of the bandgap and exciton binding energy in monolayers of WS$_2$ and WSe$_2$ for a variety of combinations with graphene and hBN layers.
The overall shift of the bandgap was found to range from 100 to 300\,meV, \violet{with an estimated theoretical limit of about 500\,meV.}
In addition, the rapid saturation of the screening effect with the thickness of the dielectric layer \arr{is} found both in theory and experiment \arr{to occur on a nanometer length scale, indicative of the ultra-short spatial range of the phenomenon and of the potential for highly local modification of the electronic structure.}
\new{In addition, we have demonstrated the flexibility of the technique by presenting a variety of material combinations including WS$_2$, WSe$_2$, graphene and h-BN in several configurations, with top and bottom alignment as well as in a "sandwich"-type structure.
It is further supplemented by theoretical calculations of predicted changes in the electronic structure for arbitrary dielectric constants in the vicinity of the monolayer, providing a map for future heterostructure design.
We thus emphasize, that by the nature of the screening effect, possible dielectric heterostructures are not restricted by the particular choice of a capping material in any specific way.}
\blue{\violet{Finally}, we demonstrated Coulomb engineering of a prototypical in-plane dielectric heterostructure, illustrating the feasibility of our approach.}
As a consequence, non-invasive patterning of dielectric layers on top of \blue{these} ultra-thin semiconductors or placing the latter on a prefabricated substrate \arr{result} in patterned circuits for the charge carriers \arr{and} will allow us to explore a variety of novel devices in the 2D plane, \violet{taking advantage of the fundamental limit for spatial variations of the electronic bandgap on the order of only a few unit cells~\cite{Rosner2016}.}
In addition to more conventional optoelectronic applications -- such as transistors, light emitters, and detectors -- becoming feasible on atomic length scales, we envision custom-made superstructures in 2D device structures, allowing for integration with photonic cavities, plasmonic nanomaterials, and single quantum emitters for the creation of new hybrid technologies. 
\new{As a consequence, the considerable strength of the Coulomb forces in atomically-thin materials is thus not only of fundamental importance, but also offers an alternative and powerful strategy towards deterministic engineering of the bandgaps in the 2D plane.}


%

\section{Methods}
Monolayer WS$_2$, \blue{mono- and few-layer graphene, and hBN} samples were \arr{produced by  mechanical exfoliation of bulk crystals (2Dsemiconductors, Inc.)} and \blue{WSe$_2$ (HQgraphene).}
The thickness of the layers was confirmed by optical contrast spectroscopy.
\rust{The heterostructures were fabricated using well-established polymer-stamp transfer techniques described in Refs.\,\onlinecite{Lee2014,Rigosi2015} for the WS$_2$ based samples and Ref.\,\onlinecite{Castellanos-Gomez2014a} for the WSe$_2$ samples.}
To study the exciton states we performed optical reflectance measurements using a tungsten-halogen white-light source. 
The light was focused to a 1\,-\,2\,$\mu$m spot on the sample for the measurements on WS$_2$, and to a 5\,-\,10\,$\mu$m spot for the measurements on WSe$_2$ due to larger sample sizes. 
The samples were kept in an optical cryostat at temperatures around 70\,K \rust{and 4\,K for the WS$_2$ and WSe$_2$ samples, respectively}.
The reflected light was spectrally resolved in a grating spectrometer and subsequently detected by a CCD.

Exciton binding energies were calculated within the Wannier-Mott model, with an exciton reduced mass obtained from DFT calculations\,\cite{Berkelbach2013}. The electron-hole screened Coulomb interaction was obtained from the quantum electrostatic heterostructure approach\,\cite{Andersen2015}. Additional details on the sample preparation, experimental procedure, and theoretical modeling can be found in the Supplementary Information.

\section{Acknowledgements}
The authors would like to thank Simone Latini and Mark S. Hybertsen for fruitful discussions.
Funding for this research was provided in part by the Center for Precision Assembly of Superstratic and Superatomic Solids, an NSF MRSEC (Award Number DMR-1420634), by FAME, one of six centers of STARnet, a Semiconductor Research Corporation program sponsored by MARCO and DARPA, as well as through the AMOS program at SLAC National Accelerator Laboratory within the Chemical Sciences, Geosciences, and Biosciences Division. 
Use of the Shared Materials Characterization Laboratory (SMCL) made possible by funding from Columbia University. 
Device fabrication was supported by the Nanoelectronics and Beyond program of the National Science Foundation (grant DMR-1124894) and the Nanoelectronics Research Initiative of the Semiconductor Research Corporation. 
An.C. acknowledges support from the Science Without Borders program of the Brazilian National Research Council (CNPq) and the Lemann Foundation.
J.Y. thanks the Kwanjeong Educational Foundation for support.  
H.M.H. and A.F.R. acknowledge funding from the National Science Foundation through the Integrated Graduate Education and Research Training Fellowship (DGE-1069240) and the Graduate Research Fellowship Program (DGE-1144155), respectively.
Al.C., T.K., P.N. and C.S. gratefully acknowledge funding from the Deutsche Forschungsgemeinschaft through the Emmy Noether Programme (CH 1672/1-1) and via GRK1570.
\noindent





\end{document}